\documentclass[journal,twoside]{IEEEtran}
\usepackage{epstopdf}
\usepackage{amsmath,amsfonts}
\usepackage{algorithmic}
\usepackage{algorithm}
\usepackage{array}
\usepackage[caption=false,font=normalsize,labelfont=sf,textfont=sf]{subfig}
\usepackage{textcomp}
\usepackage{stfloats}
\usepackage{url}
\usepackage{verbatim}
\usepackage{graphicx}
\usepackage{authblk}
\usepackage{multicol}
\usepackage{fancyhdr}
\usepackage{cite}

\providecommand{\keywords}[1]
{
  \small	
  \textbf{\textit{Keywords---}} #1
}

\fancypagestyle{mypagestyle}{%
  \fancyhf{}
  \fancyhead[OC]{Mespinosa \emph{et. al}}
  \fancyhead[EC]{Experiments in the penetration of cuboid intruders near walls into granular matter}
  \fancyfoot[C]{\thepage}%
}
\title{Experiments in the penetration of cuboid intruders near walls into granular matter}

\author[a]{M. Espinosa}
\author[a,b]{L. Alonso-LLanes} 
\author[c]{R. Herrero}
\author[a,1]{E. Altshuler}

\affil[a]{Group of Complex Systems and Statistical Physics, Physics Faculty, University of Havana, 10400 Havana, Cuba}
\affil[b]{Universit{\'e} de Strasbourg, CNRS, Institut Terre et Environnement de Strasbourg, UMR7063, 67000 Strasbourg, France}
\affil[c]{National Institute for Research in Metrology, 10200 Havana, Cuba}
\affil[1]{Corresponding author: ealtshuler{@}fisica.uh.cu}

\begin{document}

\twocolumn[
  \begin{@twocolumnfalse} 

\maketitle




\begin{abstract}

When an object penetrates into granular matter near a boundary, it experiences a horizontal repulsion due to the intruder-grain-wall interaction. Here we show experimentally that a square cuboid intruder, released from rest with no initial velocity, near a vertical wall to its left, goes through three distinct kinds of motion: it first tilts clockwise, then ``slides" away from the wall, and finally tilts counterclockwise. This dynamic highly favors both repulsion and penetration of the cuboid intruder as compared to that observed from its release farther away from the wall. 
\end{abstract}

\keywords{Granular matter, Boundary effects, Intruder penetration, Sedimentation \vspace{30px}}

\end{@twocolumnfalse}
]

\thispagestyle{empty}

Granular matter often displays unexpected phenomena due to its intrinsic discrete nature, combined with the dissipative interactions between grains \cite{le1996ticking,eggers1999sand,altshuler2003sandpile,shinbrot2004granular,martinez2007uphill,altshuler2008revolving}. For instance, the penetration of a solid intruder into a granular bed shows a non-trivial combination of solid-like and fluid-like behaviors. Over the last years, the study of low-velocity penetration of solid objects into granular matter has gained substantial momentum, especially within the Physics community \cite{uehara2003low,walsh2003morphology,boudet2006dynamics,katsuragi2007unified,goldman2008scaling,nelson2008projectile,pacheco2010cooperative,pacheco2011infinite,torres2012impact,ruiz2013penetration, brzinski2013depth,altshuler2014settling,clark2014collisional,sanchez2014note,de2016lift,viera2017note,bester2017collisional,doi:10.1080/02726351.2021.1983905}. However, the vast majority of such research focuses on the penetration away from the boundaries of the granular container: intruder-wall interactions are an almost virgin research topic, with very few reports before 2020 \cite{nelson2008projectile,katsuragi2012nonlinear}. In a more recent article, we have thoroughly studied the repulsion of cylindrical intruders near walls, and have demonstrated that the repulsion process involves rotation \cite{Diaz-Melian2020}. We have extended these studies for more than a single intruder \cite{espinosa2021intruders}.

In this article, we introduce a new subject in the filed of intruder-wall interactions. We experimentally show that a square cuboid intruder released near a vertical wall rotates around its symmetry axis, while moving away from it (tilting phase), before ``sliding" through a virtual ``slope" into the granular bed (sliding phase). The motion ends with an opposite rotation that reverts the rotated angle back to a value closer to zero (reverse tilting phase). This process occurs at a certain initial distance from the wall; for higher values, the tilted angle, the penetration depth and the horizontal displacement all sharply decrease.

\begin{figure}[!h]
\includegraphics[width=230px]{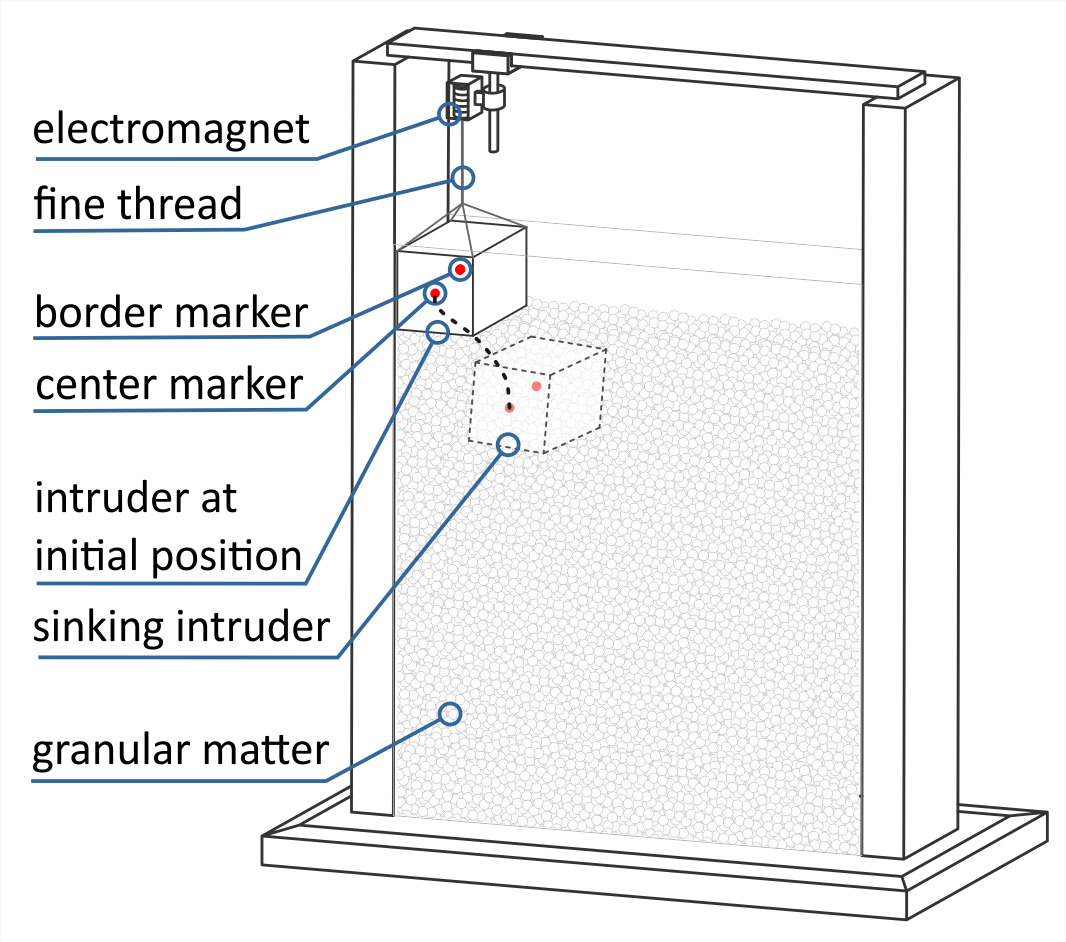}
\caption{Experimental setup. A square cuboid intruder is released into a Hele-Shaw cell as shown in the sketch.}
\label{fig:Setup}
\end{figure}

\begin{figure}[!h]
\includegraphics[width=230px]{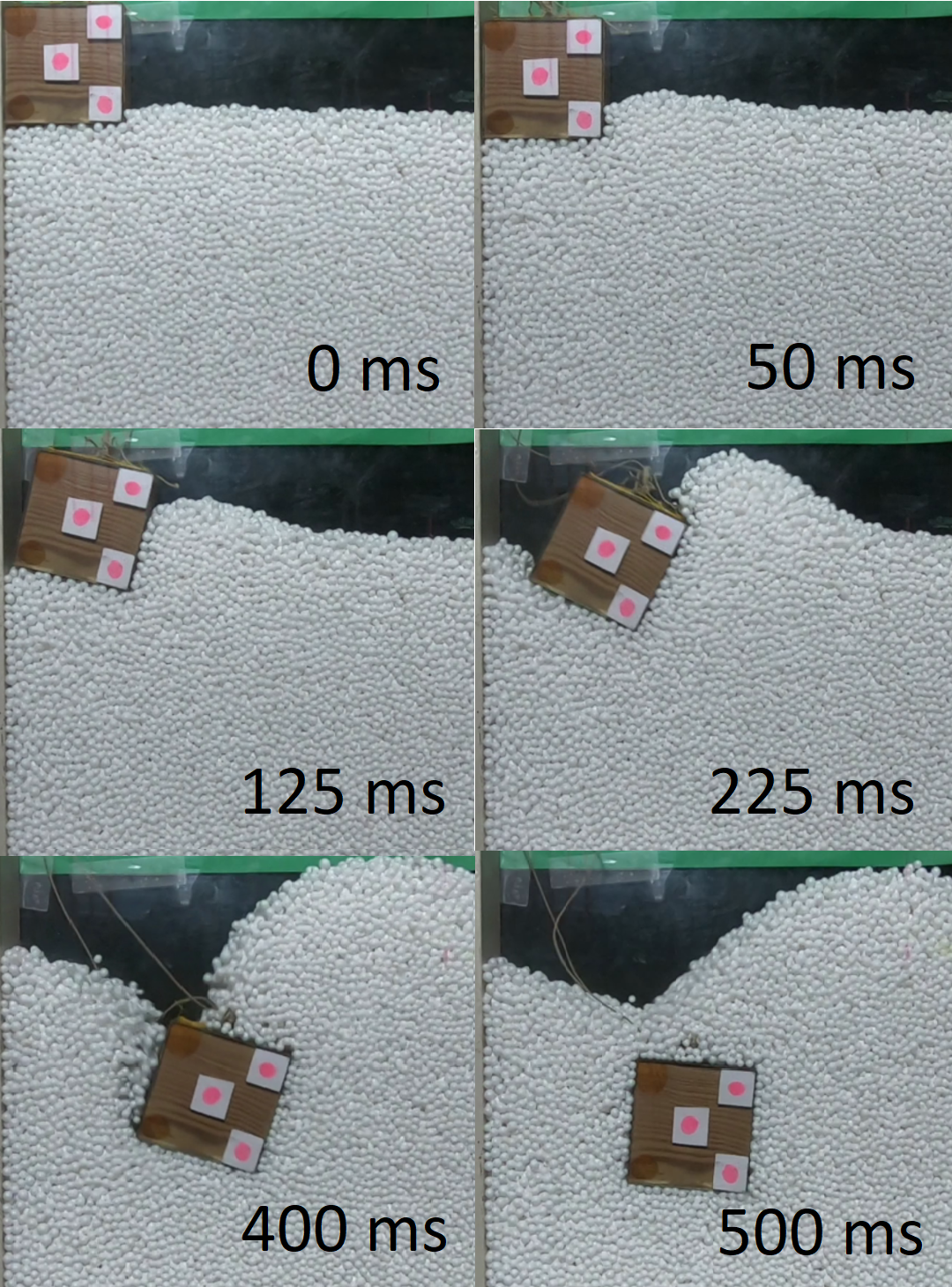}
\caption{Sequence of snapshots of the motion of the intruder released from the free granular surface touching the wall. The left wall of the cell corresponds to the left boundary of each picture. Notice the three stages of the motion: tilting, sliding and reverse tilting.}
\label{fig:Snapshots}
\end{figure}

\section*{Experimental details}
Expanded polystyrene spheres were deposited into a Hele-Shaw cell, as shown in Fig. \ref{fig:Setup} (the size distribution of the granular material and the dimensions of the cell were similar to those in \cite{Diaz-Melian2020}). A squared-face intruder of $6.8$~cm side, $5.2$ ~cm width  and $0.237$~kg mass was released from the surface of the granular bed by means of an electromagnetic device that minimized initial spurious vibrations and torques on the intruders. Initially, the intruder was gently touching the left vertical wall of the cell as illustrated in Fig. \ref{fig:Setup}. Videos of the penetration process were taken through one of the large faces of the cell using a digital camera. Three colored dots, located at the center and near the edges of one of the squared faces of the intruder, served as reference points for image analysis \cite{reyes2021yupi}, so the motion and rotation angles could be quantified.

\section*{Results and discussion}

A squared intruder released near a vertical wall shows an interesting behavior during its penetration. In Fig. \ref{fig:Snapshots} a series of snapshots of a typical experiment is illustrated. The first stage of the motion ($<50\,$ms) is a vertical plunge corresponding to an almost free fall. This is followed by a tilting phase (between $50\,$ms and $125\,$ms) with little penetration. After the intruder rotates a certain angle it ``slides" into the granular bed (between $125\,$ms and $400\,$ms) resulting in a very large lateral displacement (1.2 times the intruder size). The ``slope" over which the intruder moves is formed by the loaded force chains that form between the bottom of the intruder and the vertical wall. The motion ends with an inverse tilting (starts around $300\,$ms) that partially compensates the initial tilting, in such a way that the intruder ends with little inclination. This effect is the result of the intruder colliding into a more solid layer of the granular bed, that exerts a net torque ``correcting" its rotation.

\begin{figure}[!h]
\includegraphics[width=260px]{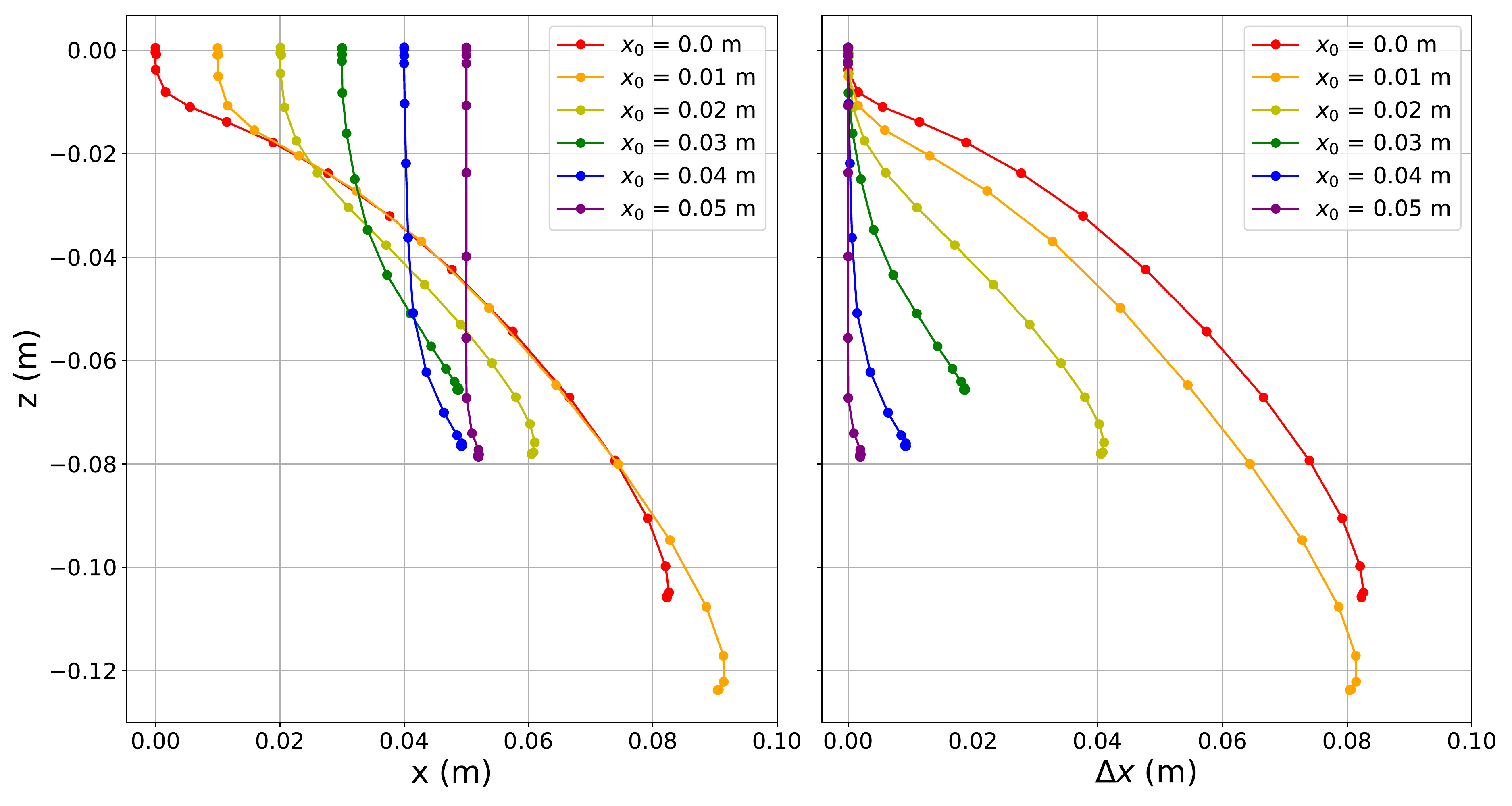}
\caption{(a) Trajectories averaged over 10 repetitions for $x_0=$ 0, 1, 2, 3, 4 and $5\,$cm. (b) Same trajectories, where the horizontal axis shows the net displacement ($\Delta x = x - x_0$).}
\label{fig:Trajectories}
\end{figure}

Fig. \ref{fig:Trajectories} indicates the trajectories of the intruder released at different initial separations ($x_0$) from the wall, resulting from the average over $10$ repetitions of the experiment. The trajectories followed by the intruders corresponding to $x_0 = 0$ and $1\,$cm are strikingly similar, suggesting the formation of the same ``slope" in both cases. The effect of the slope for higher values of $x_0$ greatly diminishes, as should be expected with the increase in the distance from the wall, resulting in less stressed force chains (``loose slope"). 

\begin{figure}[!h]
\includegraphics[width=255px]{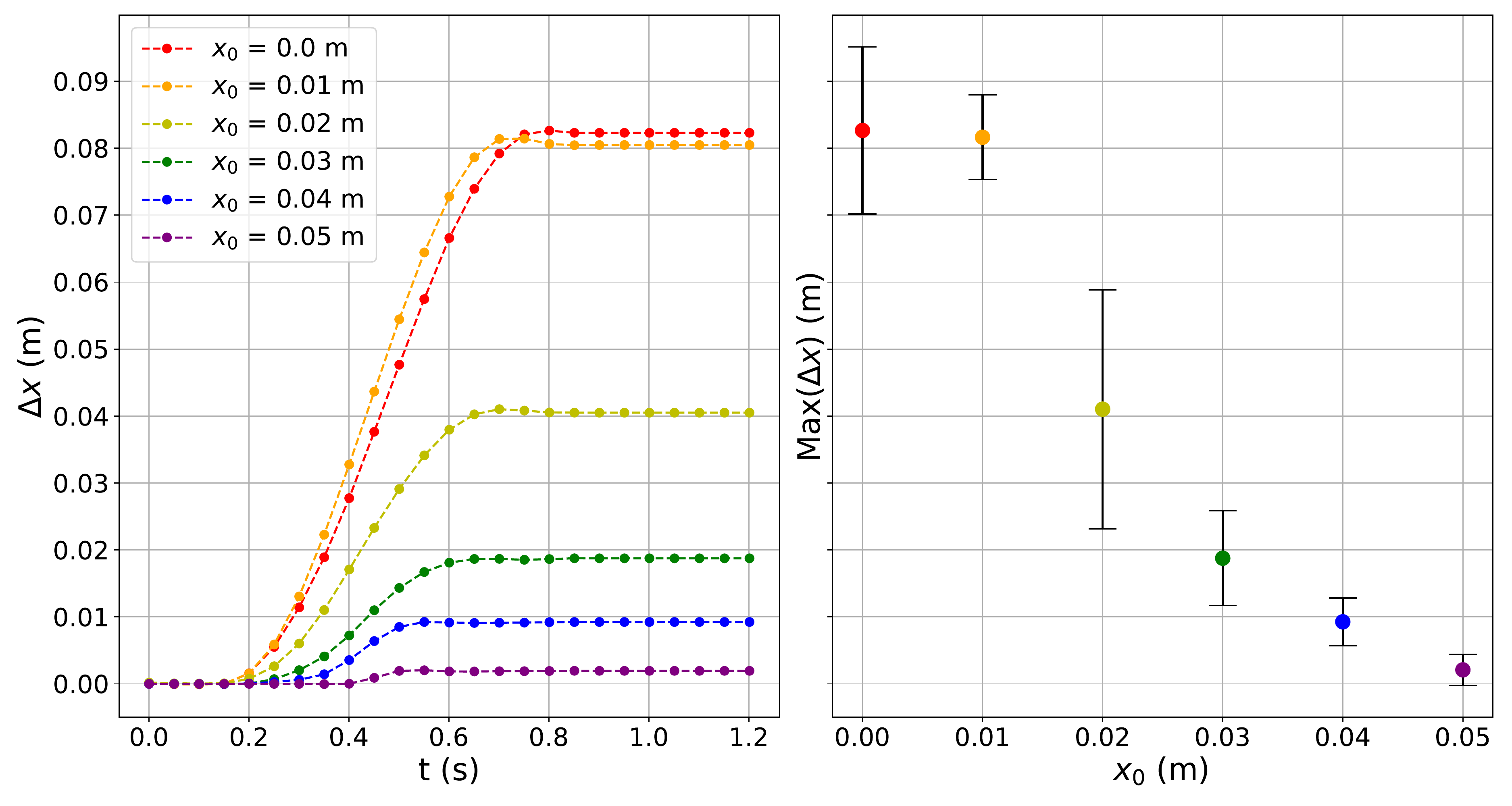}
\caption{Net horizontal repulsion. (a) Time evolution of $\Delta x = x - x_0$ of the intruder after being released for the $x_0$ values shown in Fig. \ref{fig:Trajectories}. (b) Maximum lateral displacement (error bars are the corresponding standard deviations). The color scale of both (a) and (b) is consistent with the one shown in Fig. \ref{fig:Trajectories}.}
\label{fig:deltax}
\end{figure}

Fig. \ref{fig:deltax} illustrates how much greater the lateral repulsion for $x_0 = 0$ and $1\,$cm are, compared to all other values of $x_0$. The intruders released closer to the wall slide through a more rigid slope of force chains, resulting in a very large horizontal motion. This phenomena is not observed in cylindrical intruders \cite{Diaz-Melian2020}, for which the maximum value of $\Delta x$ is much smaller (0.65 times the intruder size for a cylinder, and 1.2 times the intruder size for a square cuboid).

\begin{figure}[!h]
\includegraphics[width=255px]{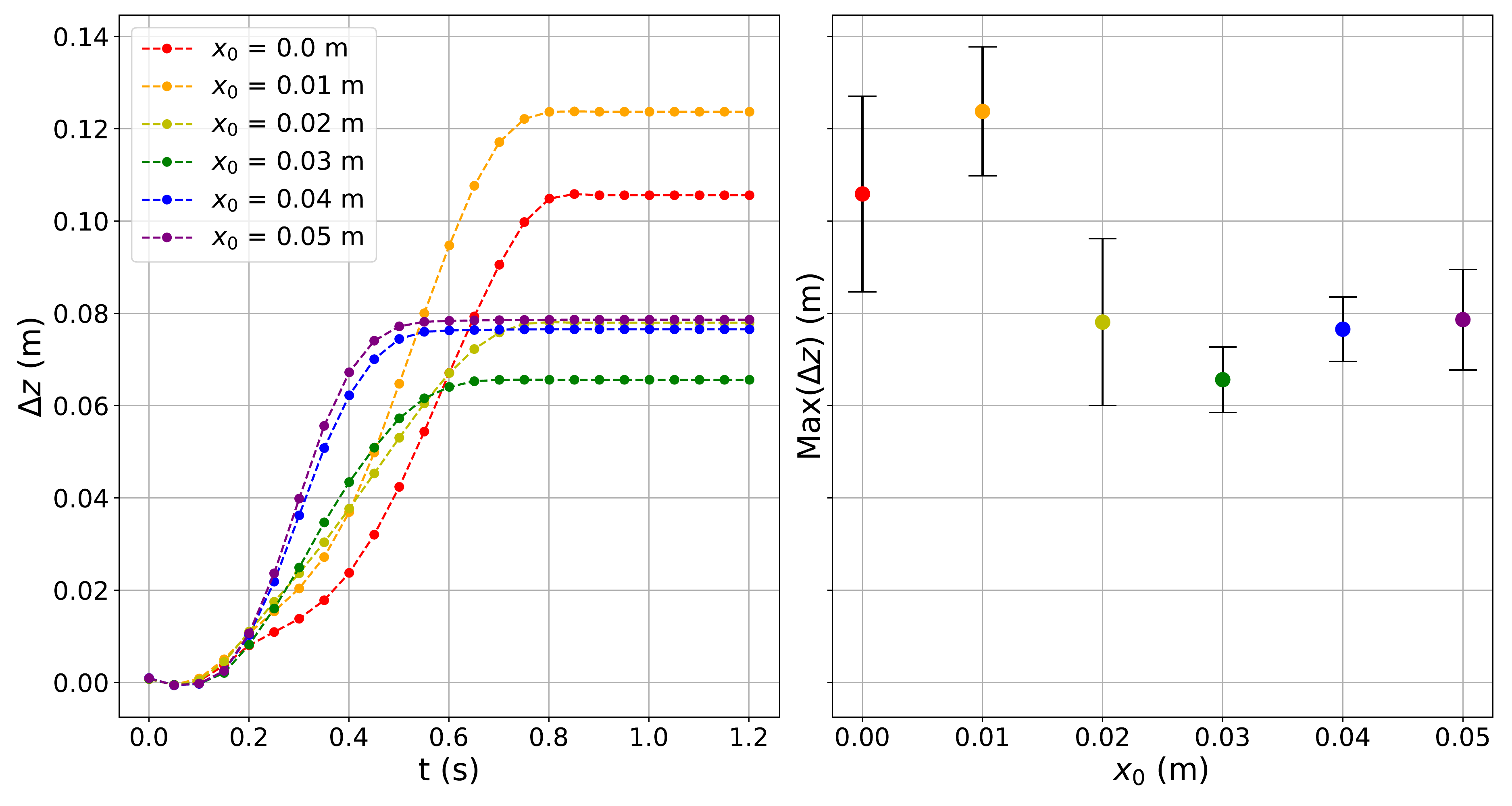}
\caption{Vertical penetration. (a) Time evolution of $\Delta z = z - z_0$ of the intruder after being released for the $x_0$ values shown in Fig. \ref{fig:Trajectories}. (b) Maximum penetration depth.}
\label{fig:deltaz}
\end{figure}

Fig. \ref{fig:deltaz} shows the vertical penetration of the intruder released from different initial position $x_0$. Here we distinguish a slightly higher penetration for $x_0=1\,$cm than for $x_0=0\,$cm, caused by the stronger force chains closer to the wall that dissipate a higher portion of the intruder's potential energy. The maximum penetration decreases for $x_0=2\,$ and $3\,$cm, due to a lesser rotation of the intruder (for $x_0 = 0$ and $1\,$cm the geometry of the rotated intruder favors the penetration). Surprisingly, the maximum value of $\Delta z$ increases at $x_0=4\,$cm, reaching a stable value (approximately the same for higher values of $x_0$). This could be caused by the reduction of the effect of the vertical wall in the stress of the force chains formed under the intruder (note that this effect was overlapped by the rotation that favored penetration for lower values of $x_0$).

\begin{figure}[!h]
\includegraphics[width=255px]{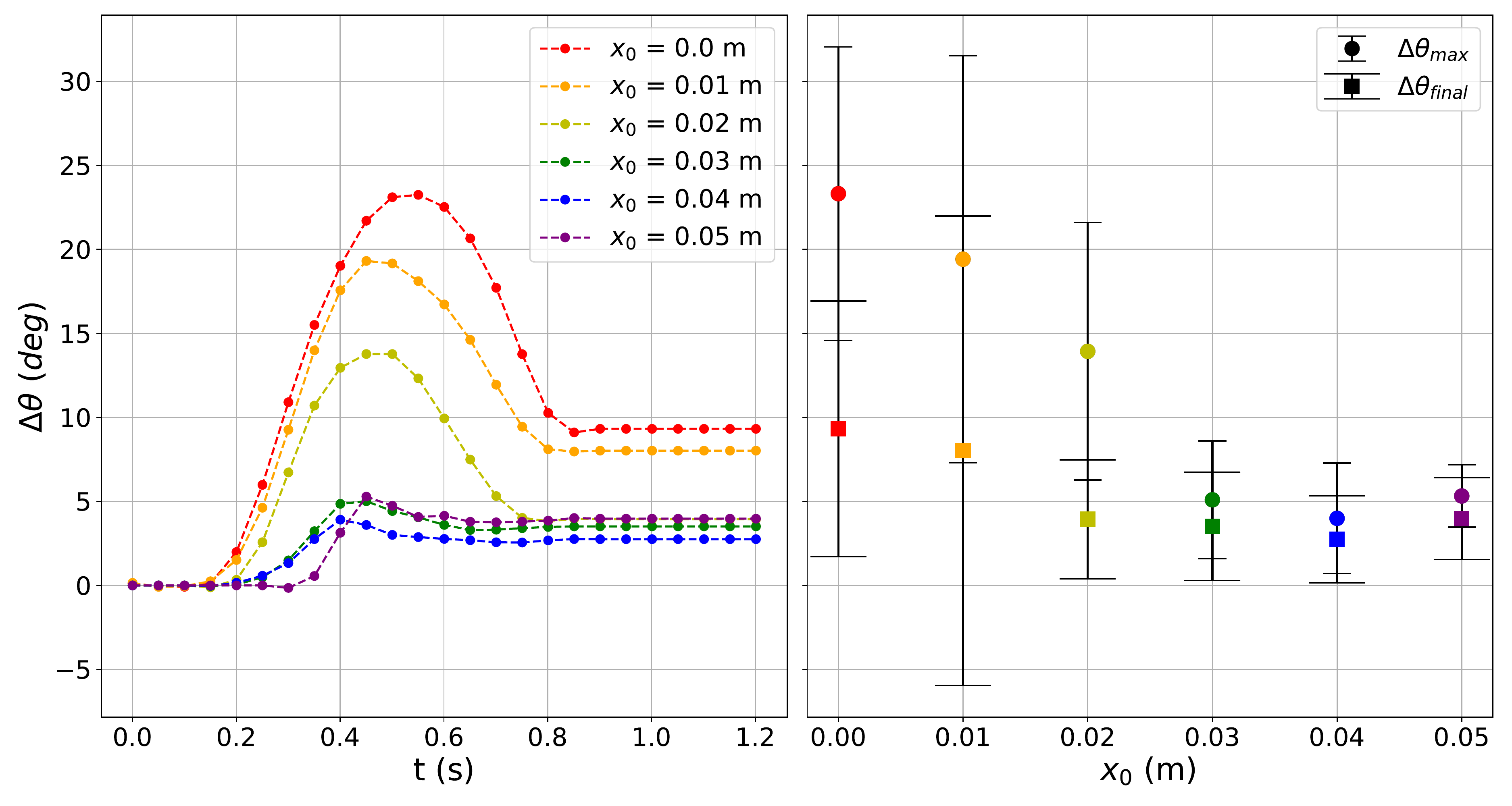}
\caption{Rotation. (a) Time evolution of $\Delta \theta = \theta - \theta_0$ of the intruder after being released. (b) Maximum and final values of $\theta$.}
\label{fig:Rotation}
\end{figure}

Fig. \ref{fig:Rotation} shows the time evolution of the rotated angle $\theta$ of the intruder released from increasing values of $x_0$. The three stages of the motion: tilting, sliding and inverse tilting can be clearly observed. As shown by the large values of the error bars for $x_0\le2\,$cm, the rotation of the intruder seems quite sensitive to the fluctuations in the configuration of the granular bed.

We are performing Discrete Element Simulations using LAMMPS (Large-scale Atomic/Molecular Massively Parallel Simulator) \cite{plimpton2007lammps} in order to reproduce experimental results, visualize the force chains, and understand the role of dissipation in the dynamics. The results will be published elsewhere.

\section*{Conclusions}

We have shown that a square cuboid intruder released near a vertical wall into a granular bed moves in three phases. Firstly, it rotates around its symmetry axis, moving away from the wall (tilting phase). Secondly, it "slides downhill" on top of a virtual “slope” into the granular bed (sliding phase). Thirdly, the motion ends with an opposite rotation that partially compensates the rotated
angle back to a value closer to zero (reverse tilting phase). However, as the cube is initially released at increasing distances from the wall, both tilting and sliding become smaller.

\section*{Acknowledgment}

We acknowledge the University of Havana's institutional project ``Granular media: creating tools for the prevention of catastrophes". The Institute ``Pedro Kourí" is thanked for allowing us using their computing cluster. E. Altshuler found inspiration in the late M. \'Alvarez-Ponte.

\bibliographystyle{unsrt}
\bibliography{biblio}

\end{document}